\documentclass[11pt]{article}
\usepackage{moriond,epsfig}

\def\ra{\rightarrow}

\def\be{\begin{equation}}
\def\ee{\end{equation}}
\def\bea{\begin{eqnarray}}
\def\eea{\end{eqnarray}}

\begin{document}
\title{MEASUREMENT OF $V_{ub}$ AT BELLE}

\author{A. LIMOSANI \\ (for the Belle collaboration)}

\address{Experimental Particle Physics, School of Physics, University
  of Melbourne, Victoria 3010, Australia} 

\maketitle\abstracts{
Results on two inclusive $V_{ub}$ measurements at Belle are presented. We
also present a new measurement of the branching fraction for $D_s\ra\phi\pi$.
A data sample of $78.1\ {\rm fb}^{-1}$ accumulated using the Belle detector at the KEKB asymmetric $e^+e^-$ collider operating at the $\Upsilon(4S)$ resonance is used. All results are preliminary.
}

\vspace{-3mm}
\section{Introduction}
\vspace{-2mm}
The large sample of
$\Upsilon(4S)$ decay events produced by KEK-B, and studied with the
Belle detector has inspired the use of new methods to measure the
Cabibbo-Kobayashi-Maskawa matrix element, $V_{ub}$.

The Belle detector is described in detail elsewhere.\cite{bellenim}  
It consists of 
a silicon vertex detector, 
a central drift chamber (CDC), 
aerogel Cerenkov counters (ACC), 
time-of-flight scintillation counters (TOF) 
and an electromagnetic calorimeter made of CsI(Tl) crystals 
enclosed in alternating layers of resistive plate chambers and iron
for $K_L/\mu$ detection and to return the flux of the 1.5 T magnetic
field.


The measurement of $|V_{ub}|$ via the study of charmless semileptonic
$B$ decays has proved difficult due to overwhelming background
from charm semileptonic $B$ decays. Traditional measurements are derived from an
examination of the charged lepton yields at the end point of the
momentum spectrum. The precision of
eventual $|V_{ub}|$ extractions are limited since at best end point measurements
access roughly 10\% of the total $B\rightarrow X_u l \nu$ phase space.
Recent work has shown methods which can measure the hadronic recoil
mass, $M_{X_u}$, and leptonic mass, $q^2\equiv(p_l+p_\nu)^2$, offer more precise $|V_{ub}|$
measurements~\cite{Bauer:2001rc}.

A model dependent $|V_{ub}|$ extraction is
possible in measurements in $B^0\ra D_s^+ \pi^-$
decays~\cite{Hayakawa:2002ss}. Both BaBar and Belle collaborations
have measured the branching fraction for $B^0\ra D_s^+ \pi^-$ decay~\cite{Aubert:2002vg}$^,\,$\cite{Krokovny:2002pe}.
The largest contribution to the systematic error is the
uncertainty in the rate of  $D_s \ra \phi \pi$. A more precise
measurement of the rate will aid any eventual $|V_{ub}|$ extraction. 

\section{Inclusive $|V_{ub}|$ measurements}




\subsection{Pseudo full reconstruction of the $\Upsilon(4S)$ via two semileptonic $B$ decays}
$\Upsilon(4S)$ decay to two $B$ mesons which both subsequently decay
semileptonically enable one to recover the $B$ direction
up to a two-fold ambiguity, if all $B$ decay products with the
exception of the neutrinos are fully reconstructed. 

\begin{figure}[b!]
  \begin{center}
    \begin{tabular}{cc}
      \includegraphics[width=0.25\columnwidth]{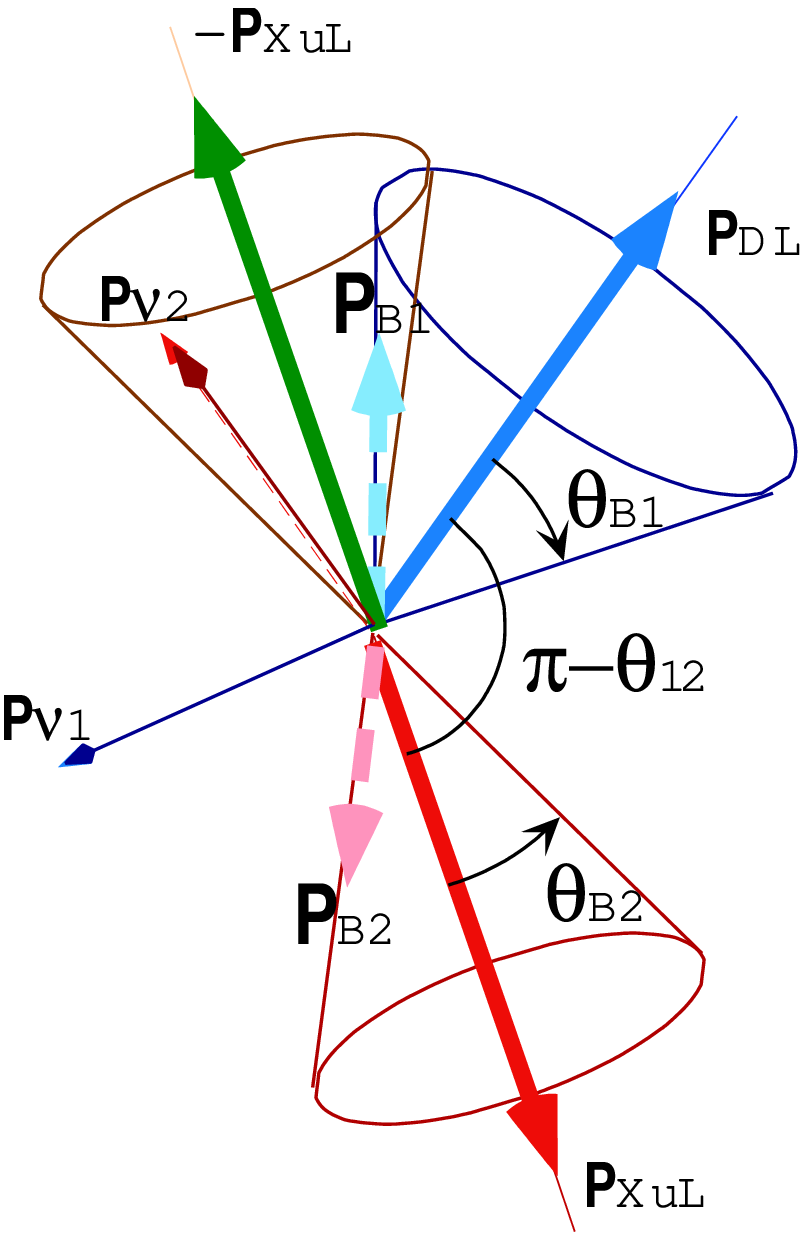} &
      \includegraphics[viewport= 0 545 242 781,clip,width=0.37\columnwidth]{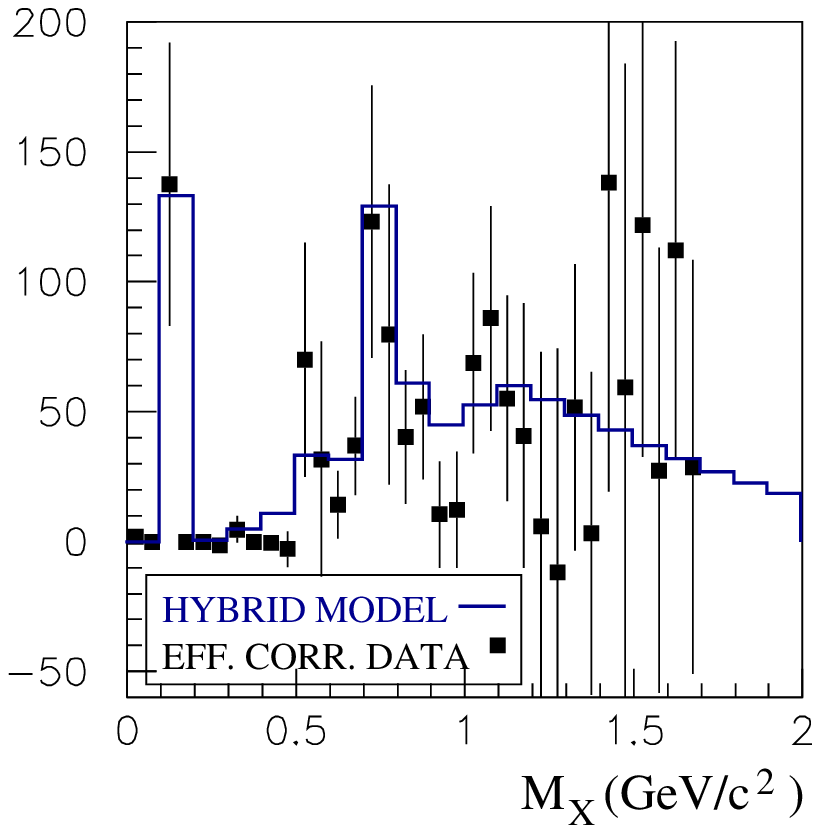} \\
    \end{tabular}
    \caption{(left) Depiction of two back to back
      semileptonic $B$ decays, intersection of the two cones
      identifies the two possible $B$ directions. (right) $M_X$ distribution of efficiency corrected data overlayed
      with Signal MC histogram (Hybrid Model -
      $\{\pi,\rho,\eta,\omega\}l \nu$ + inclusive spectrum)}
    \label{fig:pseudo_mx}
  \end{center}
\end{figure}

Signal events require a charm and charmless semileptonic $B$ decay \emph{i.e.}
$B_1\rightarrow D^{(*)} l_1 \nu$ (tag side) and $B_2 \rightarrow X_u l_2 \nu$
(signal side)($l=e,\mu$). 
Choosing for the z axis the $D^{(*)}l$ direction, one can solve for
the $B$ direction components, $(x_B,y_B,z_B)$
(see Figure~\ref{fig:pseudo_mx}),\\
\begin{eqnarray*}
x_B^2 & = & 1 - \frac{1}{\sin^2\theta_{12}}(
\cos^2 \theta_{B1} +
\cos^2 \theta_{B2} - 2 \cos \theta_{B1}
\cos\theta_{B2} \cos \theta_{12} )
\\
y_B & = & \frac{1}{\sin\theta_{12}}(\cos\theta_{B2} -
\cos\theta_{B1}\cos\theta_{12})
\\
z_B & = & \cos\theta_{B1}.
\end{eqnarray*}
For pseudo full reconstruction of the $\Upsilon(4S)$, 
we reconstruct two charged leptons($e^\pm$ or $\mu^\pm$) and a
$D^{(*)}$ meson and assign all other reconstructed particles remaining in the event
to the $X_u$ system. To suppress backgrounds, we require $p_{l_2}>
1\,\mathrm{GeV/c}$, $x_B^2 > -0.2$, no charged kaons in $X_u$,
$M_{X} < 1.5 \mathrm{GeV/c^2}$, and zero net charge for an
event. After MC background subtraction in $M_{X}$,  for the
case $X_u$ consists of at
least one $\pi^0$, we find the yield $N_{B \rightarrow X_u l \nu} =92
\pm 21$ with a reconstruction efficieny of
0.2\%, and for $X_u$ without a $\pi^0$, we find the yield
 $N_{B\rightarrow X_u l \nu} =89 \pm 19$, with a reconstruction efficieny of 0.05\%. The combined $M_X$
distribution is shown in Figure~\ref{fig:pseudo_mx}.

\subsection{Hadronic recoil mass, $M_X$, and leptonic mass, $q^2$, reconstruction with a
  Simulated Annealing technique}
This analyses employs an advanced neutrino reconstruction technique. The
requirement of only one reconstructed charged lepton($e$ or $\mu$) is
imposed on the event. The missing 4-momentum in the
event is
calculated by summing all reconstructed particle 4-momenta and
subtracting it from that of the known $\Upsilon(4S)$.  
A decision tree based on track and neutral cluster
quality is used to reduce the number of spurious particles included in
the missing momentum calculation. Neutrino reconstruction is further
improved by the reconstruction of the other $B$ decay, so called tag.
The $B_{tag}$ reconstruction is achieved through a probabalistic and
iterative process based on a Simulated Annealing technique\cite{kirk}. The technique minimises a measure,
$W$, which is calculated using $PDF$s of signal and random particle
combinations, {\emph i.e.} $W\equiv
PDF(random)/(PDF(signal)+PDF(random))$.  
A signal particle combination corresponds to the case whereby
reconstructed particles in the event are correctly assigned to either
the $B_{tag}$ or $X_u$. Minimisation of $W$ works to favour correct
assignment, and hence correct $B_{tag}$ and $X_u$ reconstruction.
The $PDF$s are functions of
the $B_{tag}$ cms momentum, $p^*_{B_{tag}}$, cms energy, $E^*_{B_{tag}}$, number of final state
decay products, $N_{B_{tag}}$, cms polar angle, $\cos\theta_{B_{tag}}^*$,
the product of $B_{tag}$ and lepton charge, $Q_{B_{tag}} \times
Q_l$, and  missing mass squared, $m^2_{miss}$. A $B_{tag}$ candidate is chosen, if  $W<0.1$,
$5.1<E_B^*<5.4\,\mathrm{GeV}$,  $0.25<p_B^*<0.42\,\mathrm{GeV/c}$, $-2<Q_B \times
Q_l < 1$, and $-0.2 < m^2_{miss} < 0.4 \, \mathrm{GeV^2/c^4}$.

\begin{figure}[b!]
  \begin{center}
  \begin{tabular}{cc}
    \includegraphics[width=0.33\columnwidth]{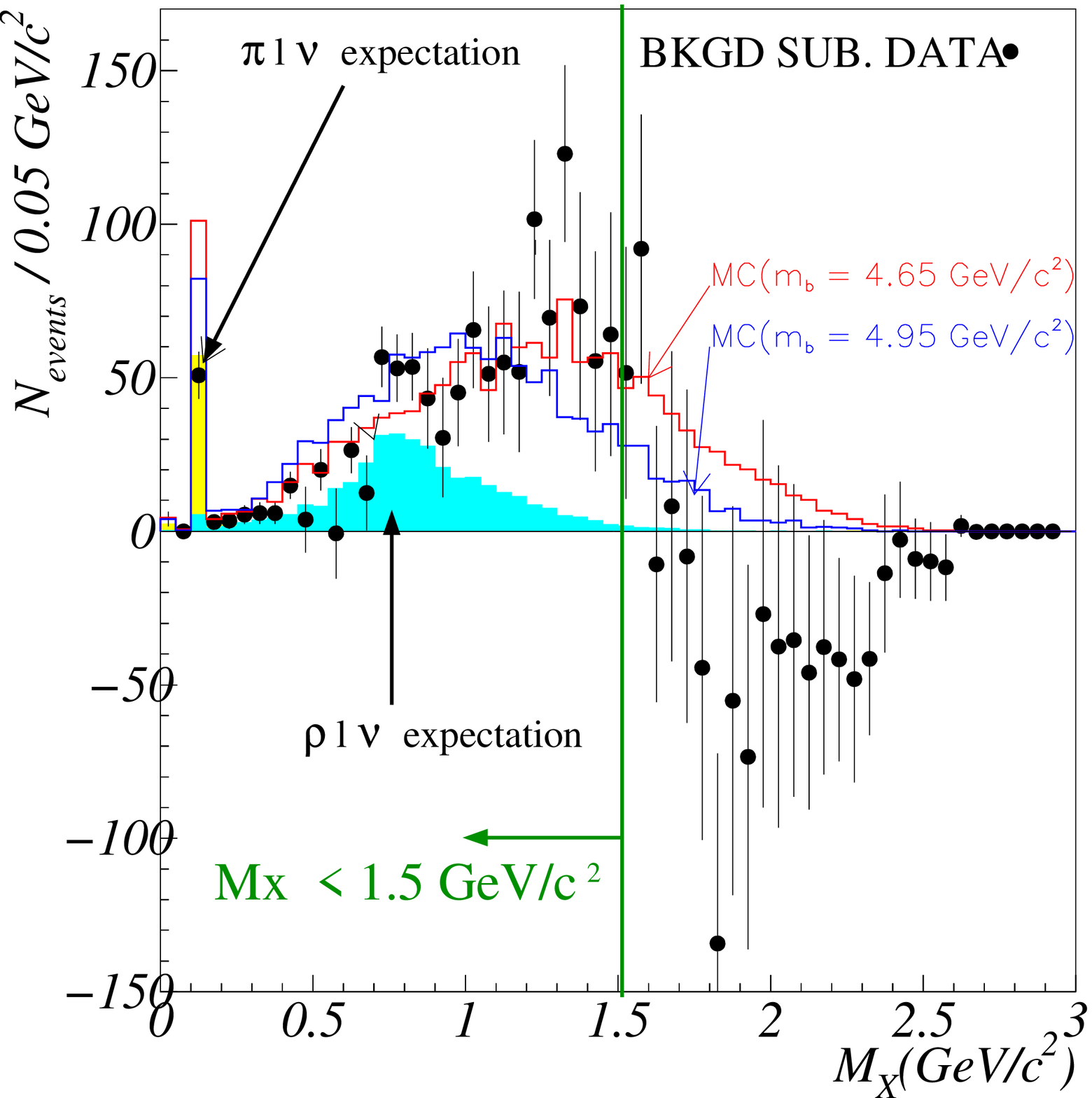}
     &
      \includegraphics[width=0.35\columnwidth]{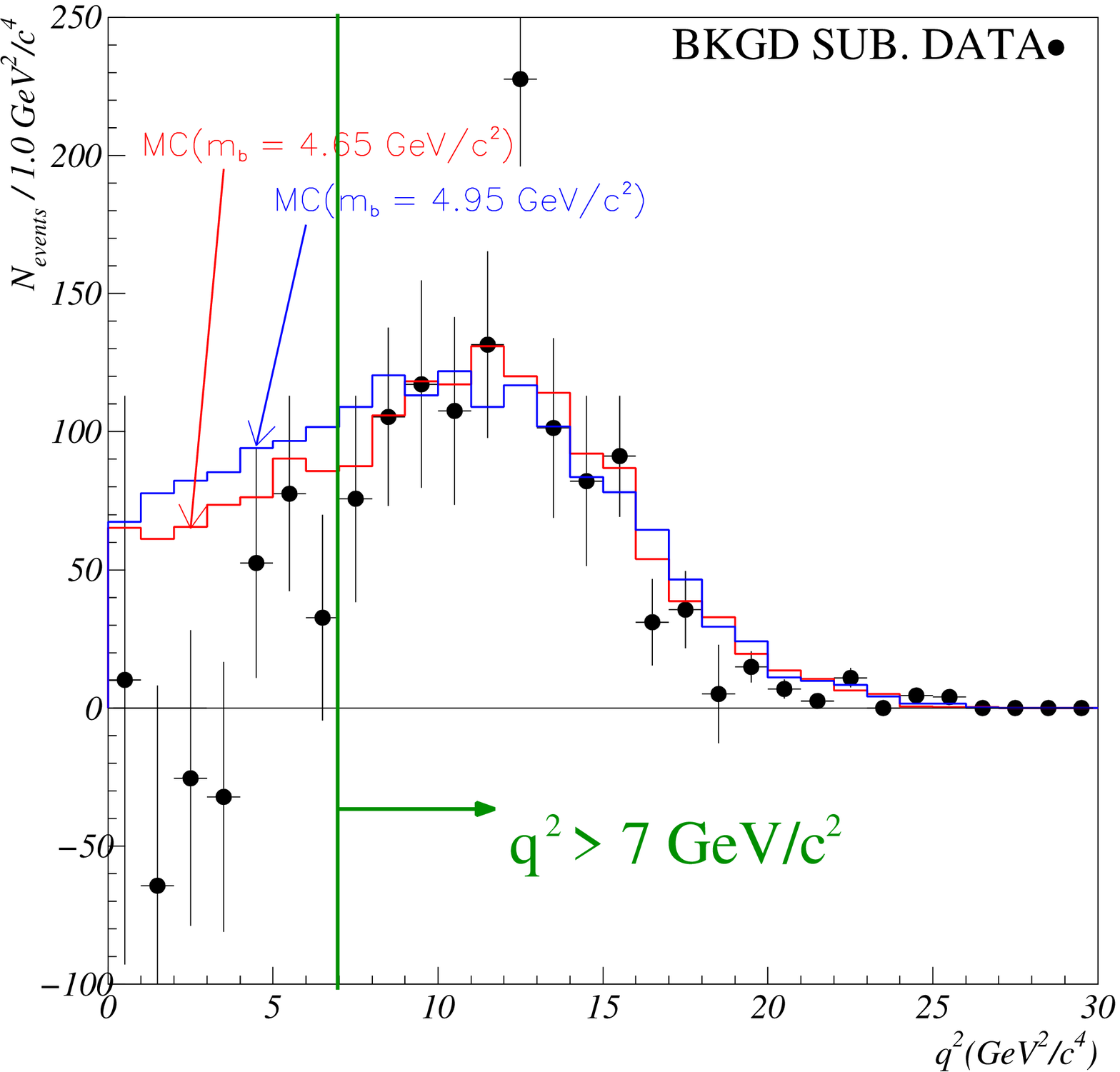} 
     \\
  \end{tabular}
  \caption{Background subtracted $M_X$ and $q^2$
    distributions overlayed with Signal MC histograms}
  \label{fig:adv_q2_mx}
  \end{center}
\end{figure}

The background from non
$\Upsilon(4S)$ events is subtracted using 8.8fb$^{-1}$ of OFF
resonance data scaled to ON resonance luminosity. 
The remaining background, estimated using
an efficiency corrected MC sample, is subtracted to yield signal $q^2$ and $M_X$ distributions as shown in Figure~\ref{fig:adv_q2_mx}.
In the signal region $M_{X}<1.5\,\mathrm{GeV/c}$ and $q^2 > 7\,\mathrm{GeV/c}$, we find
$N_{B\ra X_u l \nu } = 1270$ with a signal to noise ratio of 0.31.

\subsection{$\mathcal{B}(B\rightarrow X_u l \nu )$ and $|V_{ub}|$ measurement}
The inclusive rate for charmless semileptonic $B$ decay for the two
analyses discussed are shown in Table~\ref{tab:vubresults}. The
uncertainty in the shape and level of the charm semileptonic
background estimated from MC samples is quoted as the $b\ra c$ error. The uncertainty in the
extrapolation from the partial to the full phase space for $B \ra X_u
l \nu$ is quoted as the $b \ra u$ error. The value of $|V_{ub}|$ is
calculated according to the PDG 2002
formulation~\cite{Bigi:1997fj}$^-$\cite{Ligeti:1999yc}, the
uncertainty in its use is quoted as the 'theory' error.

\begin{table}[h]
  \begin{center}
    \begin{tabular}{|l|cccccc|}
      \hline
      & \multicolumn{6}{c|}{Pseudo full reconstruction of $\Upsilon(4S)$} \\
      & & statistical & systematic & $b \ra c$ &  $b \ra u$ &
      theory \\
      $\mathcal{B}(B\rightarrow X_u l \nu )(\times 10^{-3})$ & $2.62$ &
      $\pm 0.63$ & $\pm 0.23 $ & $\pm 0.05 $ & $\pm 0.41$ & \\
      $|V_{ub}|(\times 10^{-3})$ & $5.00$ &  $\pm 0.60$ & $\pm 0.23$
      & $ \pm 0.05 $ & $\pm 0.39$  & $\pm 0.36$  \\
      \hline
      & \multicolumn{6}{c|}{$q^2$ and $M_X$ reconstruction through Simulated Annealing} \\
      & & statistical & systematic & $b \ra c$ &  $b \ra u$ &
      theory \\
      $\mathcal{B}(B\rightarrow X_u l \nu )(\times 10^{-3})$ & $1.64$ &
      $\pm 0.14$ &  $\pm 0.36 $ & $\pm 0.28 $ & $\pm 0.22$ & \\
      $|V_{ub}|(\times 10^{-3})$ & $3.96$ &  $\pm 0.17$ & $\pm 0.44$
      &   $\pm 0.34$ &  $\pm 0.26$  & $\pm 0.29$ \\
      \hline
    \end{tabular}
    \caption{The inclusive branching fraction and $|V_{ub}|$
      measurements}
    \label{tab:vubresults}
  \end{center}
\end{table}  
\vspace{-7mm}
\section{Measurement of the branching fraction $D_s^+ \ra \phi \pi^+$}
We measure the branching fraction, $\mathcal{B}(D_s^+ \ra \phi\pi^+)$
by calculating an efficiency corrected ratio of full to
partially reconstructed $B\ra D^{*-}_sD^{*+}$ decays\cite{Artuso:1996xr}. In full reconstruction,
the $D_s$ is reconstructed through the $\phi\pi$ decay channel,
whereas in partial reconstruction the $D_s$ is not reconstructed,
rather a soft photon is used to tag the  $D^{*-}_s$ decay. 

\begin{figure}[t!]
\begin{center}
  \label{fig:ds}
  \begin{tabular}{cc}
     \includegraphics[width=0.33\columnwidth]{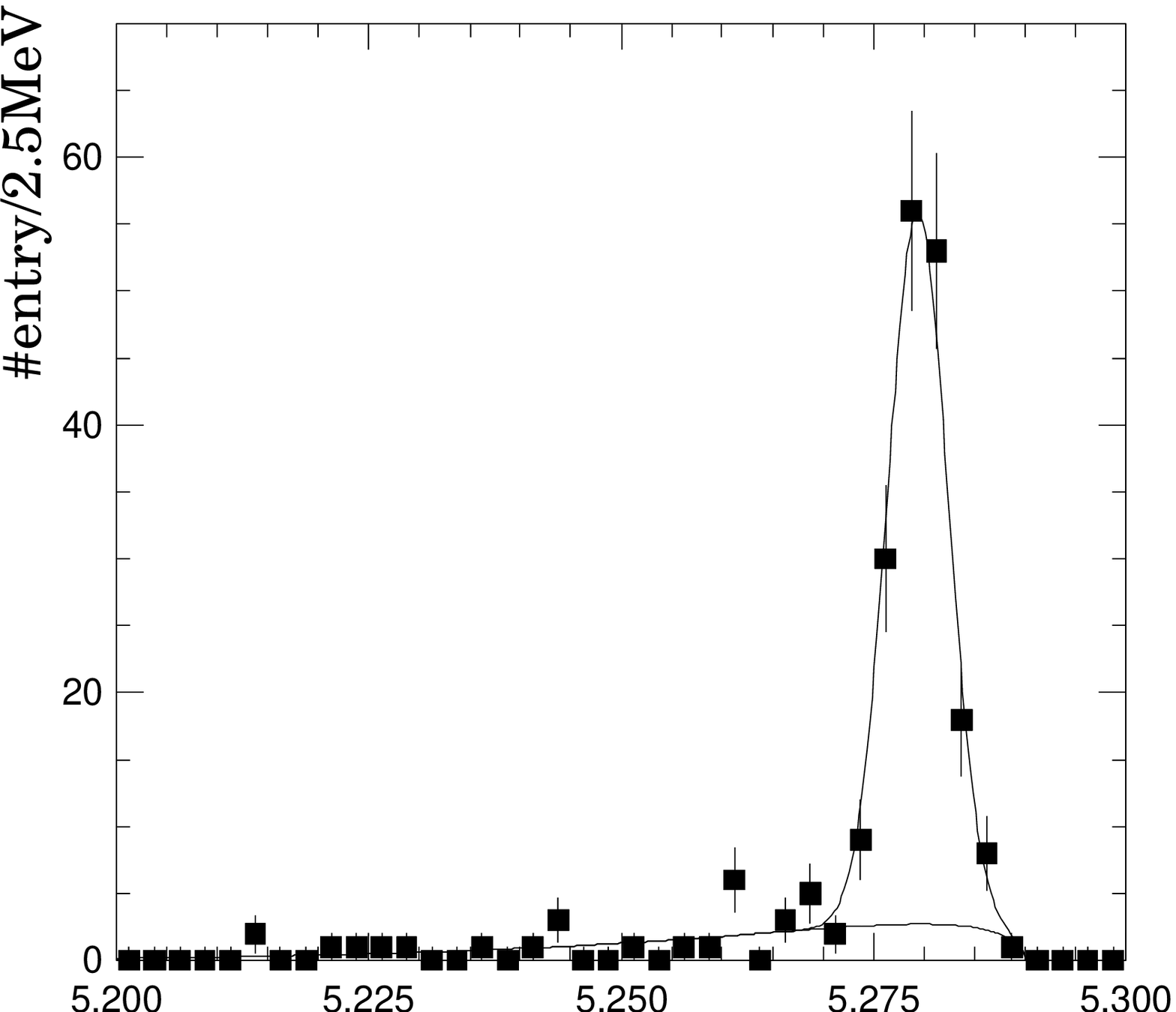} &
     \includegraphics[width=0.33\columnwidth]{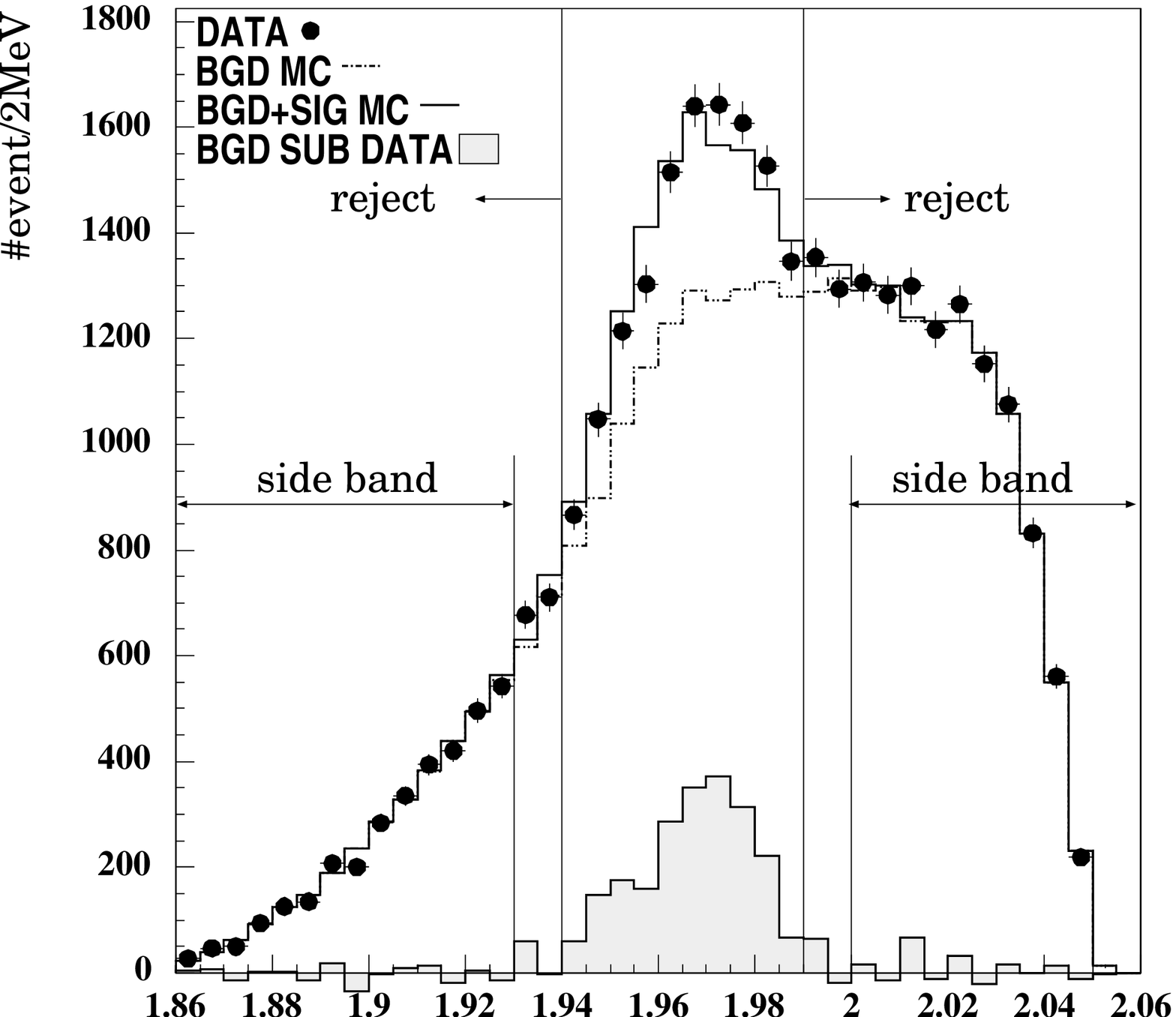} \\
     $M_{bc}$ [GeV/c$^2$] & $M_{\mathrm{miss}}$ [GeV/c$^2$]  
  \end{tabular}
  \caption{(left) Beam constrained mass for fully reconstructed $B$
    candidates and (right) Missing $D_s$ mass distribution for
    partially reconstructed $B$ candidates}
\end{center}
\end{figure}

Fully reconstructed $B$ candidates are selected based on the beam
constrained mass, $m_{bc}\equiv\sqrt{E_{beam}^2 - |\vec{p}_{B^0}|^2 }$,
which must be within 5.27 and 5.29 GeV/c$^2$, and energy difference,
$\Delta E \equiv E_{beam}-E_{B^0}$, which must lie between -0.05 and
0.04 GeV (all quantities are calculated in the $\Upsilon(4S)$
rest frame).
For events with multiple $B$
candidates, the one with the smallest $\Delta E$ is kept. The yield of
full reconstructed $B\ra D^{*-}_sD^{*+}$, $N_{full}=159.8 \pm 14.1$,  is
calculated from a fit to the $M_{bc}$ distribution with a Gaussian for
the signal component and an ARGUS function~\cite{Albrecht:1990am} for the non $\Upsilon(4S)$ (continuum)
background component as shown in Figure~\ref{fig:ds}.

In partial reconstruction, a $D^{*-}$ candidate is fully reconstructed
along with a soft photon,
$\gamma$. The signal is extracted from the missing
$D_s$ mass, distribution, $M_{\mathrm{miss}}$, calculated using the
assumption 
$\cos\theta_{B^0\gamma} \simeq -\cos\theta_{B^0 D^{*-}}$, since the
soft photon and $D^{*-}$ are almost back to back, and using the
nominal $B$ energy in the c.m. frame along with $D^{*-}$ and $\gamma$ four-momentum. 
The yield of
partially reconstructed $B \ra D^{*-}_sD^{*+}$ decays, $N_{\mathrm{partial}}=2150.0 \pm 139.5$ is got after a
subtraction of MC modeled background normalised to the sideband, see
Figure~\ref{fig:ds}. We measure 
$
     \mathcal{B}(D_s^+ \rightarrow \phi \pi^+) =  (3.72 \pm
     0.39^{+0.47}_{-0.39})\times 10^{-2}
$

\section{Conclusion}
We have discussed new inclusive $V_{ub}$
analyses whose methods are novel and original, and the results of which
compare well with previous measurements. Aided by our large data
sample we have also reported an
improved measurement of the $D_s\ra \phi\pi$ branching
fraction. 
%
%

\end{document}